\documentclass[aps,onecolumn,showpacs]{revtex4}
\usepackage{caption2}
\usepackage{float}
\usepackage{amsmath}
\usepackage{amssymb}
\usepackage{graphicx}
\usepackage{subfigure}

\begin{document}

\title{Interactions of solitons with complex defects in Bragg gratings}
\author{Peter Y. P. Chen$^{1}$, Boris A. Malomed$^{2}$, and Pak L. Chu$^{3}$}
\address{$^{1}$School of Mechanical and Manufacturing Engineering,
University of New South Wales, Sydney 2052, Australia\\
$^{2}$Department of Interdisciplinary Studies, School of
Electrical Engineering, Faculty of Engineering, Tel Aviv
University, Tel Aviv 69979, Israel\\
$^{3}$Cotco Holdings Limited, Photonics Centre, Hong Kong Science
Park, Hong Kong}

\begin{abstract}
We examine collisions of moving solitons in a fiber Bragg grating with a
\textit{triplet} composed of two closely set repulsive defects of the
grating and an attractive one inserted between them. A \textit{doublet}
(dipole), consisting of attractive and repulsive defects with a small
distance between them, is considered too. Systematic simulations demonstrate
that the triplet provides for superior results, as concerns the capture of a
free pulse and creation of a \textit{standing optical soliton}, in
comparison with recently studied traps formed by single and paired defects,
as well as the doublet: 2/3 of the energy of the incident soliton can be
captured when its velocity attains half the light speed in the fiber (the
case most relevant to the experiment), and the captured soliton quickly
relaxes to a stationary state. A subsequent collision between another free
soliton and the pinned one is examined too, demonstrating that the impinging
soliton always bounces back, while the pinned one either remains in the same
state, or is kicked out forward, depending on the collision velocity and
phase shift between the solitons.
\end{abstract}

\maketitle

\section{Introduction}

It is well known that the interplay of the linear transmission spectrum of a
fiber Bragg gratings (FBG), which includes the bandgap, and material Kerr
nonlinearity gives rise to robust optical pulses, which are often called gap
solitons, or, more generally, Bragg-grating solitons. The solitons in FBGs
were first predicted theoretically \cite{Aceves} (see also review \cite%
{Sterke}), and then created in the experiment \cite{experiment}; later,
solitons where also created in a photonic grating, induced in an ordinary
fiber as a superposition of two mutually coherent pump beams \cite%
{photoinduced}. The stability of solitons in the standard FBG model was
investigated in detail by means of approximate semi-analytical \cite{Rich}
and accurate numerical \cite{stability} methods.

A well-known feature of the FBG model is that the soliton may have zero
velocity, which suggests that fiber gratings may be appropriate media for
the creation of stable pulses of \textit{slow}/\textit{standing light},
which is a problem of current interest \cite{slow}, including the quest of
slow optical solitons \cite{slowsoliton}. In the first experiments in the
FBG, the smallest velocity of the soliton was $\simeq 0.5c_{0}$, with $c_{0}$
the speed of light in the fiber. Recently, an essentially new experimental
result has been reported \cite{recent}, with the soliton's velocity reduced
to $0.16c_{0}$, which, quite feasibly, can be made still smaller by means of
the technique applied in Ref. \cite{recent}, \textit{viz}., the use of an
\textit{apodized} FBG, with the grating strength gradually increasing along
the fiber (theoretically, the use of an apodized FBG for the retardation of
solitons and, eventually, bringing them to a halt was elaborated in detail
in Ref. \cite{weJMO}).

Collisions of moving FBG solitons with local defects of the grating has also
been the subject of several theoretical works \cite{Weinstein}-\cite{we}. A
general objective of these studies was to predict a possibility of trapping
a moving (relatively fast) soliton by a defect, and thus transforming it
into a stable \textit{pinned optical soliton} \cite{Weinstein,weJOSAB}. Such
an outcome may be of considerable interest, as, besides being a challenging
problem to fundamental nonlinear optics and nonlinear-wave dynamics in
general, it may find applications to the design of all-optical memory and
logic elements, the trapped soliton playing the role of a data bit. In
addition, the theoretical analysis has demonstrated that the attenuation of
the pinned soliton due to the loss in the FBG\ material may be compensated
by a localized gain, applied at the same position where the pinning defect
was created \cite{wePRE}. Besides the soliton proper, one can also consider
the interaction of a more general localized wave packet with FBG defects
\cite{Martijn}. Similar interaction were examined in models of photonic
crystals, which combine a Bragg grating and layers of resonantly absorbing
atoms (see a review of the topic in Ref. \cite{Kozhekin}); in that case, a
soliton may interact with an intrinsic defect of the photonic crystal \cite%
{Mantsyzov}.

An attractive local defect in the FBG may be realized as a local suppression
of the grating. A repulsive defect is possible too, in the form of a short
segment with enhanced Bragg reflectivity. In Refs. \cite{Weinstein,weJOSAB},
it was concluded that three different outcomes of the collision of a free
soliton with an attractive defect are possible, \textit{viz}., transmission
(i.e., passage of the soliton through the defect), capture, and splitting of
the incident soliton into three pulses -- transmitted, trapped, and
reflected ones. Besides that, in the case when the incident soliton is
``heavy" (i.e. intrinsically unstable \cite{Rich,stability}), almost all the
energy may go into the reflected pulse, making the attractive defect to look
as a repulsive one \cite{weJOSAB}.

As concerns the primary goal, i.e., efficient capture of a relatively fast
free soliton, it was demonstrated in Ref. \cite{we} that better results
might be provided not by a simple attractive defect, but rather by
structures built as pairs of attractive defects separated by some distance,
or, still more efficiently, by \textit{cavities}, in the form of a pair
repulsive local defects. While a single repulsive inhomogeneity cannot trap
anything, the cavity has a strong potential for that (the trapped soliton
then performs shuttle oscillations in the cavity). It was demonstrated that
the paired defects, unlike the single attractive one, give rise to a
well-defined region in the respective parameter space where the capture of
the fast soliton is especially efficacious (up to velocities of the incident
solitons $\simeq 0.35c_{0}$ and $\simeq 0.45c_{0}$, in the cases of the
attractive and repulsive pairs, respectively).

The subject of the present work is to extend the analysis to collisions of
solitons in the FBG with more complex (but still compact and quite simple,
as concerns the fabrication) defects, built as a \textit{doublet} (alias
``dipole") and, chiefly, \textit{triplet}, i.e., respectively, a pair of
attractive and repulsive defects with a relatively small distance between
them, or two repulsive defects with an attractive one placed between them.
We conclude that the triplet is \emph{superior} to all other trapping
configurations: it provides for the capture of $2/3$ of the energy of the
incident soliton at velocities attaining $0.5c_{0}$, and, which is quite
important too, the trapped soliton quickly relaxes into a stationary state.
The ``quality" of the trapped soliton is attested to by its collision with a
fresh soliton impinging upon the trapped one: while in other settings
(including the newly introduced doublet) the outcome of the collision is
actually unpredictable, as it strongly depends on the intrinsic state of the
trapped soliton, that changes in time, for the soliton trapped by the
triplet the outcome is reproducible. The fresh soliton always bounces back,
while the pinned one either gets released, or remains pinned. These results
may be useful in terms of the above-mentioned soliton-based memory and logic
elements.

\section{The model}

Following Refs. \cite{Weinstein,weJOSAB} and \cite{we}, the FBG model with
imperfections is based on the normalized coupled-mode equations for
amplitudes of the right- and left-traveling electromagnetic waves, $u$ and $v
$:

\begin{eqnarray}
i\frac{\partial u}{\partial t}+i\frac{\partial u}{\partial x}+v+\left(
|u|^{2}/2+|v|^{2}\right) u &=&\kappa f(x)v,  \label{utux} \\
i\frac{\partial v}{\partial t}-i\frac{\partial v}{\partial x}+u+\left(
|v|^{2}/2+|u|^{2}\right) v &=&\kappa f(x)u,  \label{vtvx}
\end{eqnarray}%
where $x$ and $t$ are the coordinate along the fiber and time. The
cross-phase-modulation coefficient, group velocity of the carrier waves, and
Bragg reflectivity in the uniform grating are scaled to be $1$. Function $%
f(x)$ in these equations describes the defect of the grating, with strength $%
\kappa $. Each individual defect is realized as a rectangular configuration,
with $f(x)=1/\Delta x$ within a segment of length $\Delta x=$ \ $0.469$, and
$f(x)=0$ elsewhere, $\kappa >0$ and $\kappa <0$ corresponding to the
attractive and repulsive defects, respectively. The rectangular shape of the
defect approximates the $\delta $-function profile assumed in previous
studies \cite{Weinstein,weJOSAB,we}. In physical units, $\Delta x=1$
typically corresponds to length $\sim 1$ mm \cite{Sterke}, while the actual
size of the solitons created in the experiment may be $\lesssim 1$ cm \cite%
{experiment,recent}, hence this approximation of $\delta (x)$ seems
reasonable.

In this work, we deal with $f(x)$ corresponding to a dipole (doublet) set,
composed of attractive and repulsive rectangular defects with a small
distance between their edges, $\Delta x/4=0.117$, and (chiefly) a triplet
set, consisting of two repulsive defects with an attractive one between
them, with the same separation between adjacent defects (it was checked
that, for the doublet and triplet alike, this small separation provides for
optimum results, as concerns the capture of fast solitons).

The attractive defect is realized as a narrow segment of the FBG with
suppressed grating. Therefore, as the full local reflectivity in Eqs. (\ref%
{utux}) and (\ref{vtvx}) cannot be negative, only $\kappa f(x)\leq 1$, i.e.,
values $\kappa \leq \Delta x\equiv $ \ $0.469$, are meaningful. Unlike that,
$\kappa <0$, which corresponds to a local enhancement of the grating, is not
subject to a specific limitation on $|\kappa |$ \cite{we} (however, the form
of the coupled-mode equations for a strong grating may alter \cite{deep}).

Solitons moving at velocity $c$ (with $|c|<1$) in the uniform FBG are given
by the well-known solutions to Eqs. (\ref{utux}) and (\ref{vtvx}) with $%
\kappa =0$ \cite{Aceves}:
\begin{eqnarray}
\left\{ u,v\right\} _{\mathrm{sol}} &=&\pm e^{i\Phi }\sqrt{\frac{2\left(
1\pm c\right) }{3-c^{2}}}\left( 1-c^{2}\right) ^{1/4}\left\{ W,W^{\ast
}\right\} ,  \notag \\
W(X) &=&\frac{\sin \,\theta }{\cosh \left( X\sin \,\theta -i\theta /2\right)
},  \label{movsol} \\
\,\Phi &=&\frac{4c}{3-c^{2}}\mathrm{\tan }^{-1}\left[ \tanh \left( X\sin
\,\theta \right) \tan \frac{\theta }{2}\right] -T\cos \theta ,  \notag
\end{eqnarray}%
\noindent where $X\equiv \left( x-ct\right) /\sqrt{1-c^{2}}$, $T\equiv
\left( t-cx\right) /\sqrt{1-c^{2}}$, and the soliton's amplitude, $\theta $,
takes values $0<\theta <\pi $. Soliton family (\ref{movsol}) is stable for $%
\theta \leq \theta _{\mathrm{cr}}(c)$, where $\theta _{\mathrm{cr}%
}(c=0)\approx 1.01\left( \pi /2\right) $ \cite{Rich,stability}, and $\theta
_{\mathrm{cr}}$ very weakly depends on $c$, up to $c=1$ \cite{stability}. We
will consider solitons with $c<0$; accordingly, the dipole set is built as a
right attractive defect followed by a left repulsive one.

An exact solution is also available for a quiescent soliton pinned by the
defect with $f(x)=\delta (x)$ \cite{weJOSAB},%
\begin{equation}
\left\{ u,v\right\} =\pm \sqrt{\frac{2}{3}}e^{-it\cos \theta }\frac{\sin
\theta }{\cosh \left[ \left( x+a\,\mathrm{sgn\,}x\right) \sin \mathrm{\,}%
\theta \mp i\theta /2\right] },  \label{pinned}
\end{equation}%
with $\tanh \left( a\sin \theta \right) \,=\tanh \left( \kappa /2\right)
\cot (\theta /2)$. This solution exists for $2\tan ^{-1}\left( \tanh
\left\vert \kappa /2\right\vert \right) <\theta <\pi $.

Equations (\ref{utux}) and (\ref{vtvx}), with $\kappa \neq 0$, conserve the
corresponding Hamiltonian, and the norm (frequently called \textit{energy}
in optics), $E\equiv \int_{-\infty }^{-\infty }\left( \left\vert
u\right\vert ^{2}+\left\vert v\right\vert ^{2}\right) dx$. For the exact
soliton solution (\ref{movsol}), $E_{\mathrm{sol}}\equiv 8\theta \left(
1-c^{2}\right) /\left( 3-c^{2}\right) $, while, for solution (\ref{pinned}),
$E=(8/3)\left[ \theta -\left( \pi /2-\sin ^{-1}\left( \mathrm{\ sech\,}%
\kappa \right) \right) \mathrm{sgn}\,\kappa \right] $ \cite{weJOSAB}. If $%
|\kappa |$ is small enough, the defect with $f(x)=\delta (x)$ creates an
effective potential for a slowly moving soliton, $U(\xi )=-(8/3)\kappa
\left( \sin ^{2}\theta \right) /\left[ \cosh \left( 2\xi \sin \theta \right)
+\cos \theta \right] $, where $\xi $ is the coordinate of the soliton's
center, and its effective mass is $M=(8/9)\left( 7\sin \theta -4\theta \cos
\theta \right) $ \cite{weJMO}. In accordance with what was said above, this
potential is attractive and repulsive for $\kappa >0$ and $\kappa <0$,
respectively.

\section{Capture of a free soliton}

\begin{figure}[th]
\subfigure[]{\includegraphics[width=3in]{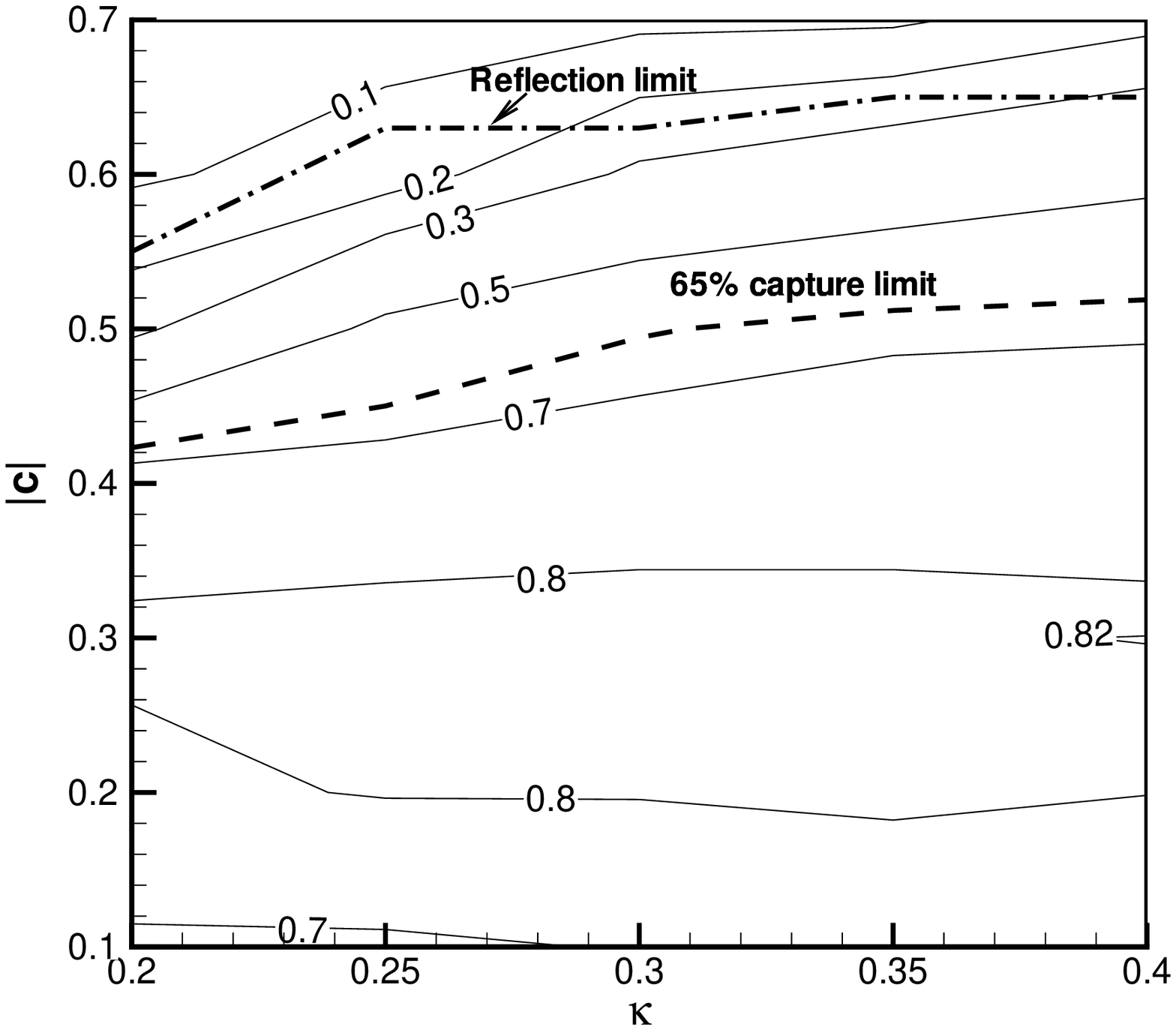}} \subfigure[]{%
\includegraphics[width=3in]{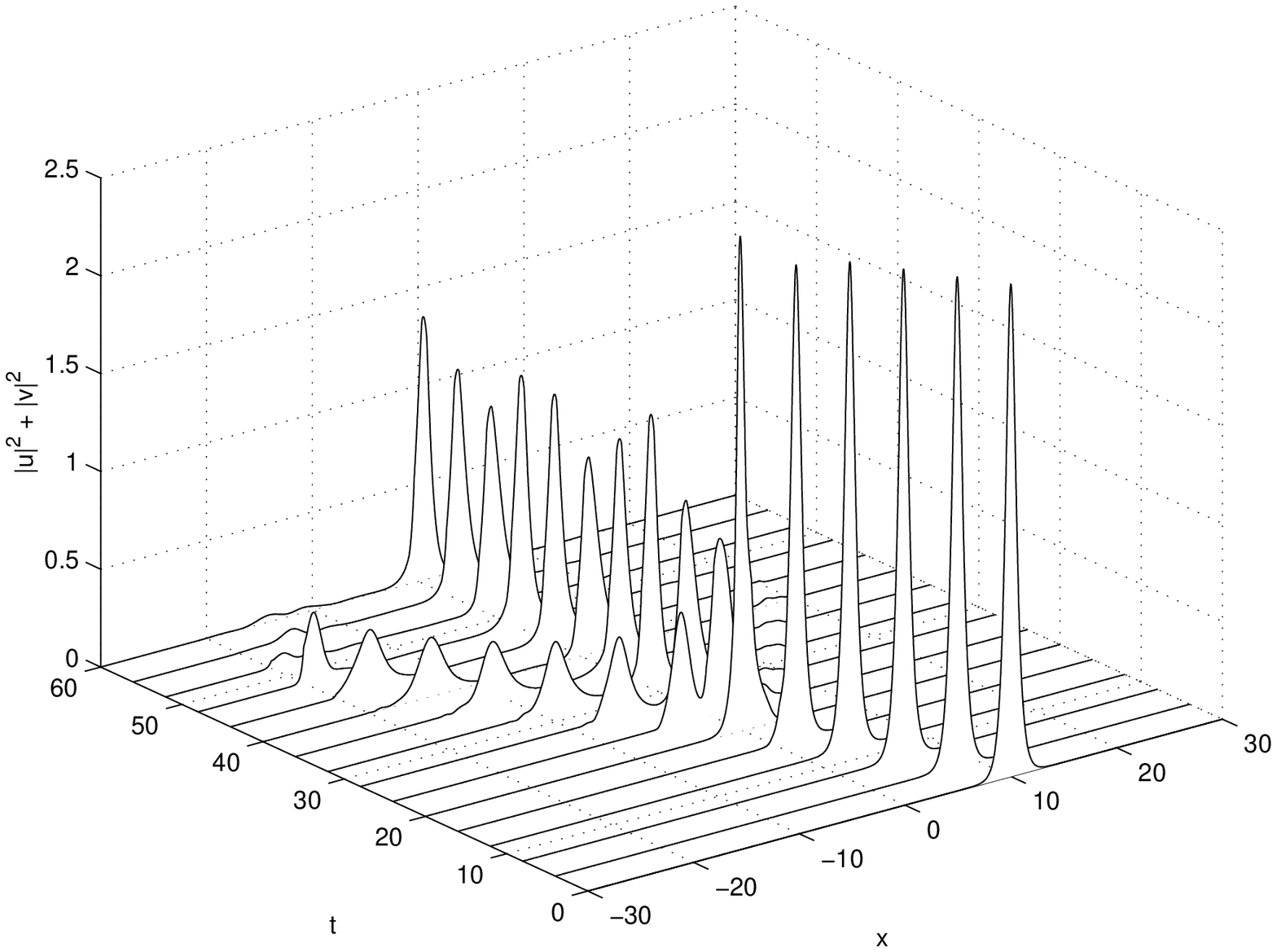}}
\caption{} \label{dipole}
\end{figure}

Numerical simulations of the collision of a free soliton with the
dipole set (the attractive defect followed by the repulsive one)
give rise to results summarized in Fig. \ref{dipole}(a), which
displays the share of the soliton's energy trapped after the
collision, the remaining energy being chiefly reflected (at small
$|c|$) or transmitted (at larger $|c|$) in the form of small
pulses, as shown in Fig. \ref{dipole}(b). We define the effective
capture of the soliton as an outcome of the collision with the
share of the trapped energy $\geq 65\%$. Figure \ref{dipole}(a)
demonstrates that the capture by the doublet, below the ``$65\%$
capture limit" boundary, is possible for $|c|\leq 0.52$, which is
a better result
than the capture limit for a pair of repulsive defects, $|c|\leq 0.42$ \cite%
{we} (which, in turn, was better than for single or paired attractive
defects \cite{weJOSAB,we}). However, Fig. \ref{dipole}(b) clearly shows that
the doublet traps the pulse in a strongly perturbed state; for this reason,
the outcome of its subsequent collision with another free soliton is
practically unpredictable, as it strongly depends on the intrinsic state of
the trapped pulse at the moment of the new collision.

The triplet produces essentially better results. In fact, it may be
considered as a generalization of the cavity formed by a pair of far
separated repulsive defects, which was introduced in Ref. \cite{we};
however, the small separation between them in the triplet, and the
attractive defect inserted in the middle, help to capture more energy from
the incident soliton, and shape the trapped energy into a stationary pulse.
A typical example of the capture by the triplet is displayed in Fig. \ref%
{3Dpictures}(a). A noteworthy peculiarity of this figure, in comparison with
the capture of solitons by single or paired defects \cite{weJOSAB,we}, as
well as by the dipole [see Fig. \ref{dipole}(b)], is that the trapped
soliton immediately settles down into a stationary state, without any
conspicuous intrinsic dynamics.
\begin{figure}[tbp]
\subfigure[]{\includegraphics[width=3in]{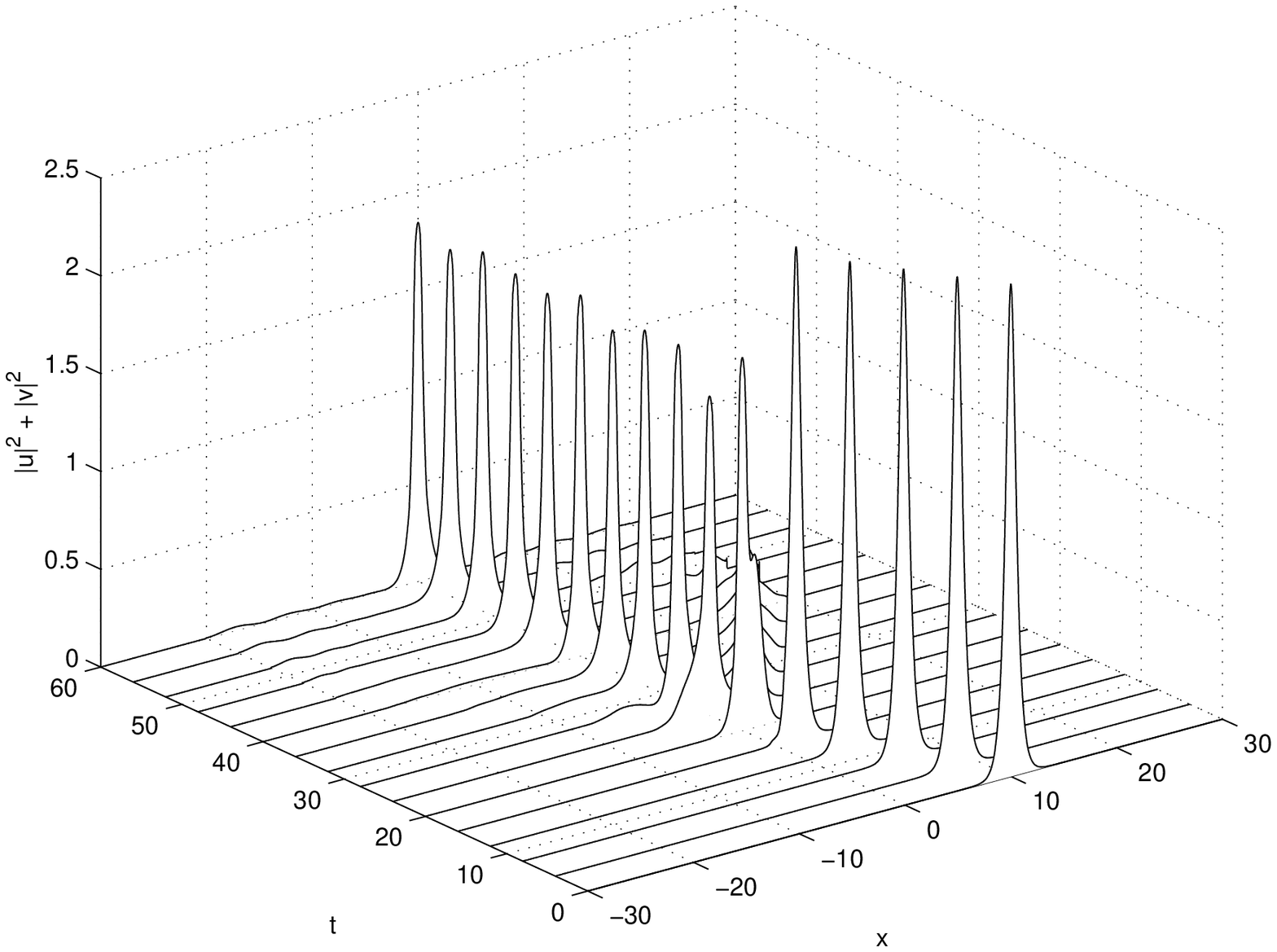}} \subfigure[]{%
\includegraphics[width=3in]{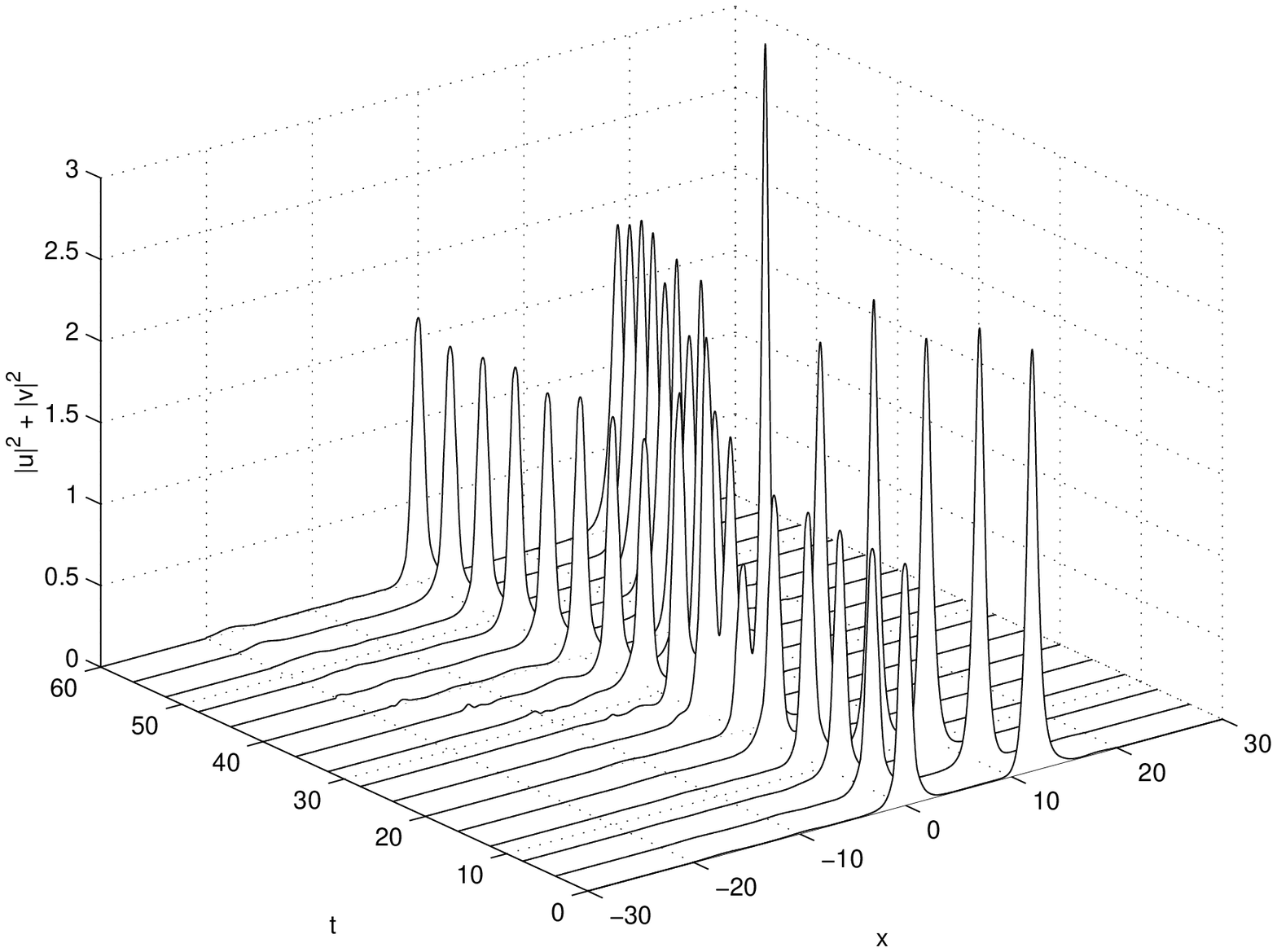}}
\caption{} \label{3Dpictures}
\end{figure}

\begin{figure}[tbp]
\subfigure[]{\includegraphics[width=3in]{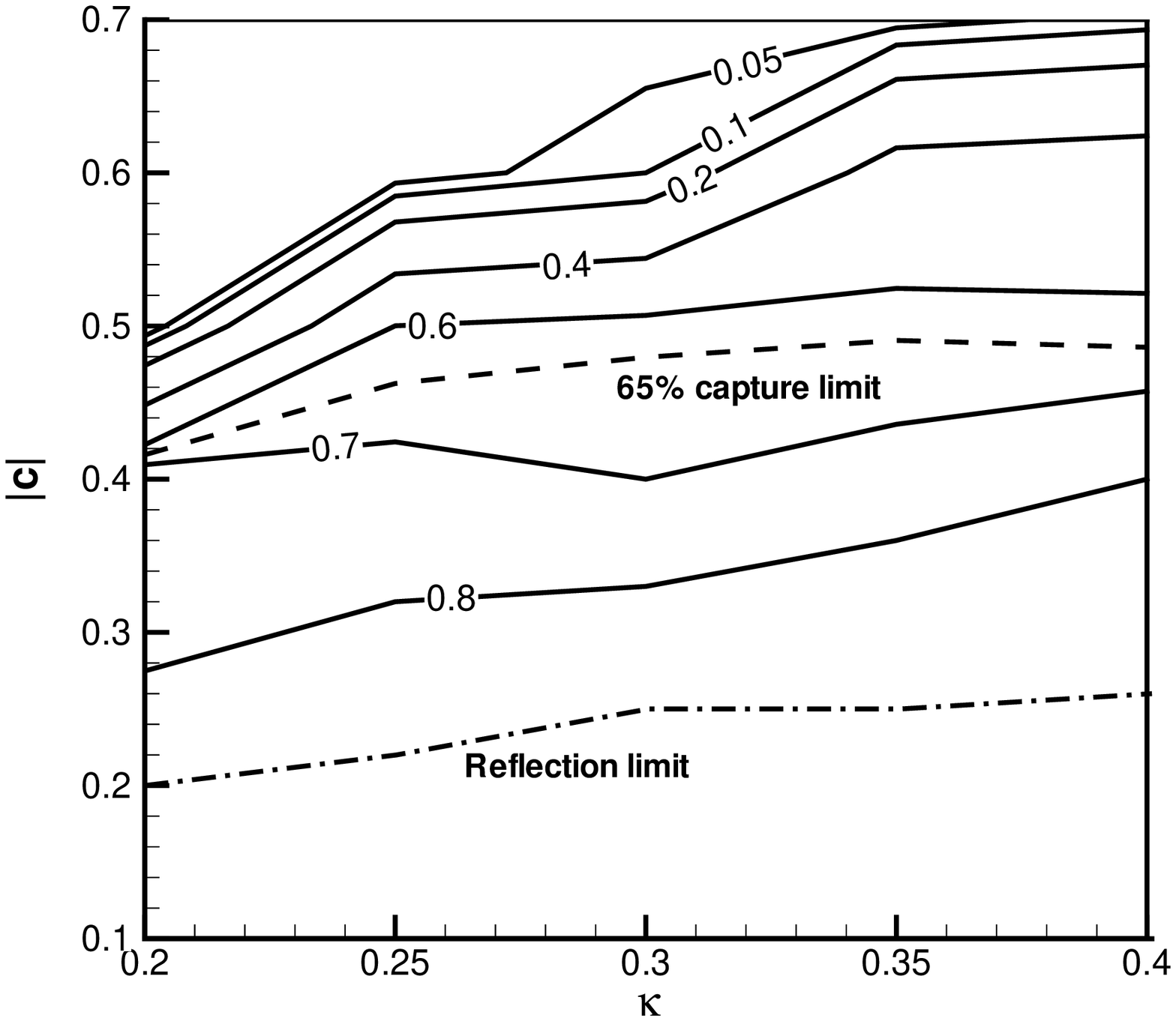}} \subfigure[]{%
\includegraphics[width=3in]{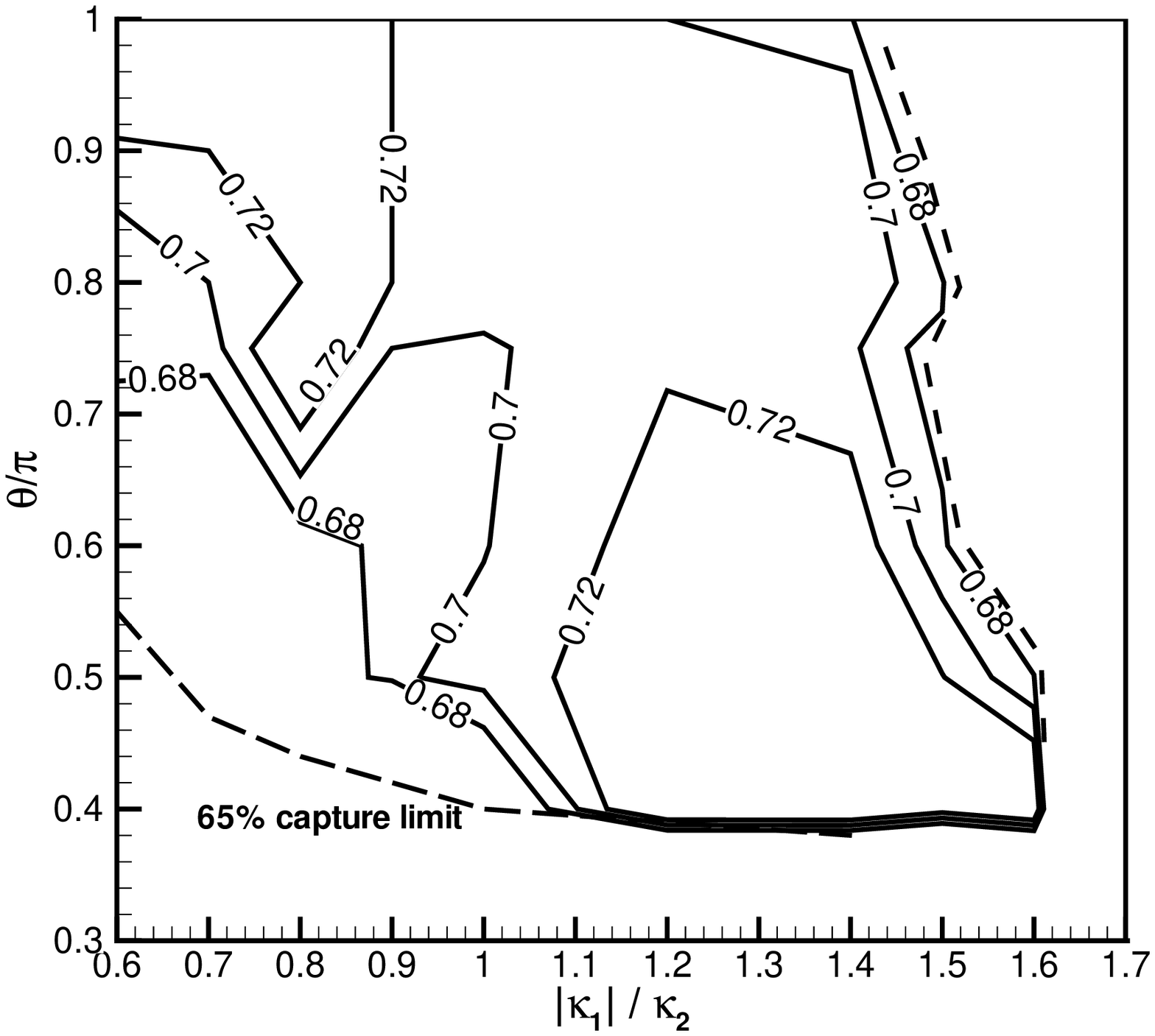}}
\caption{} \label{triplet}
\end{figure}

In a broad parameter range, the efficiency of the capture by the
triplet is characterized by plots shown in Fig. \ref{triplet}. The
capture limit is roughly the same as provided by the dipole, cf.
Fig. \ref{dipole}(a), but the quality of the trapping is much
better in the present case, as the trapped soliton is a stationary
one, as demonstrated above. It is noteworthy that the capture
efficiency is not sensitive to details of the triplet's
structures, as seen in Fig. \ref{triplet}(b), which displays the
share of the trapped energy versus the relative strength of the
repulsive and attractive components of the triplet. The same
figure demonstrates that ``heavy" incident solitons, with $\theta
>\pi /2$, which are intrinsically unstable \cite{Rich,stability},
may also be used for the creation of stable trapped solitons, if
the collision happens before the ``heavy" soliton develops its
instability. Another interesting feature of Fig. \ref{triplet}(b)
is that there is a well-defined lower threshold for the amplitude
of the free soliton, $\theta _{\min }\approx 0.4\pi $, which is
necessary for the efficient capture.

\section{Collisions between free and pinned solitons}

Collisions between bound and free solitons were not studied before in the
FBG model, even for a single attractive defect. Examined were collisions of
two free solitons, against the backdrop of the single \cite{wePRE} or paired
\cite{we} defects. Sundry outcomes were found, including the capture of both
solitons, which might merge into a single one \cite{we} (collisions between
free solitons in the uniform-FBG model were studied in detail in Ref. \cite%
{weCollision}).

As mentioned above, the outcome of the collision between a free soliton and
one pinned by the dipole defect is very irregular, if the pinned soliton
itself was generated by trapping an originally free one by the empty dipole,
as, in that case, the pinned soliton performs persistent intrinsic
vibrations, and the outcome depends on the phase of the vibrations at the
moment of the collision with the incident soliton. We have checked that
essentially the same is observed if the first soliton was originally trapped
by a single attractive defect. Obviously, this irregularity would impede the
use of the collisions in applications.

\begin{figure}[tbp]
\subfigure[]{\includegraphics[width=3in]{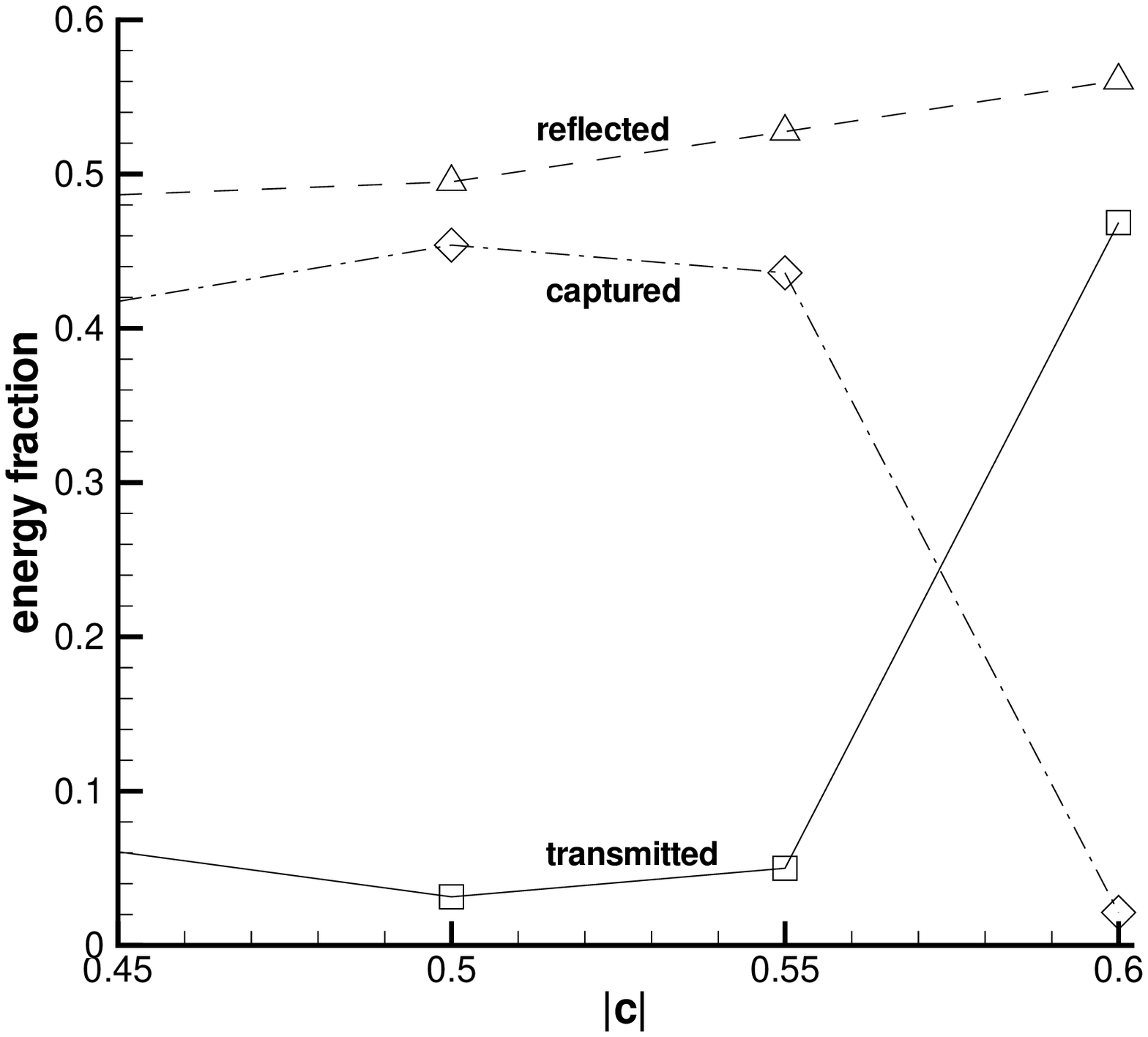}} \subfigure[]{%
\includegraphics[width=3in]{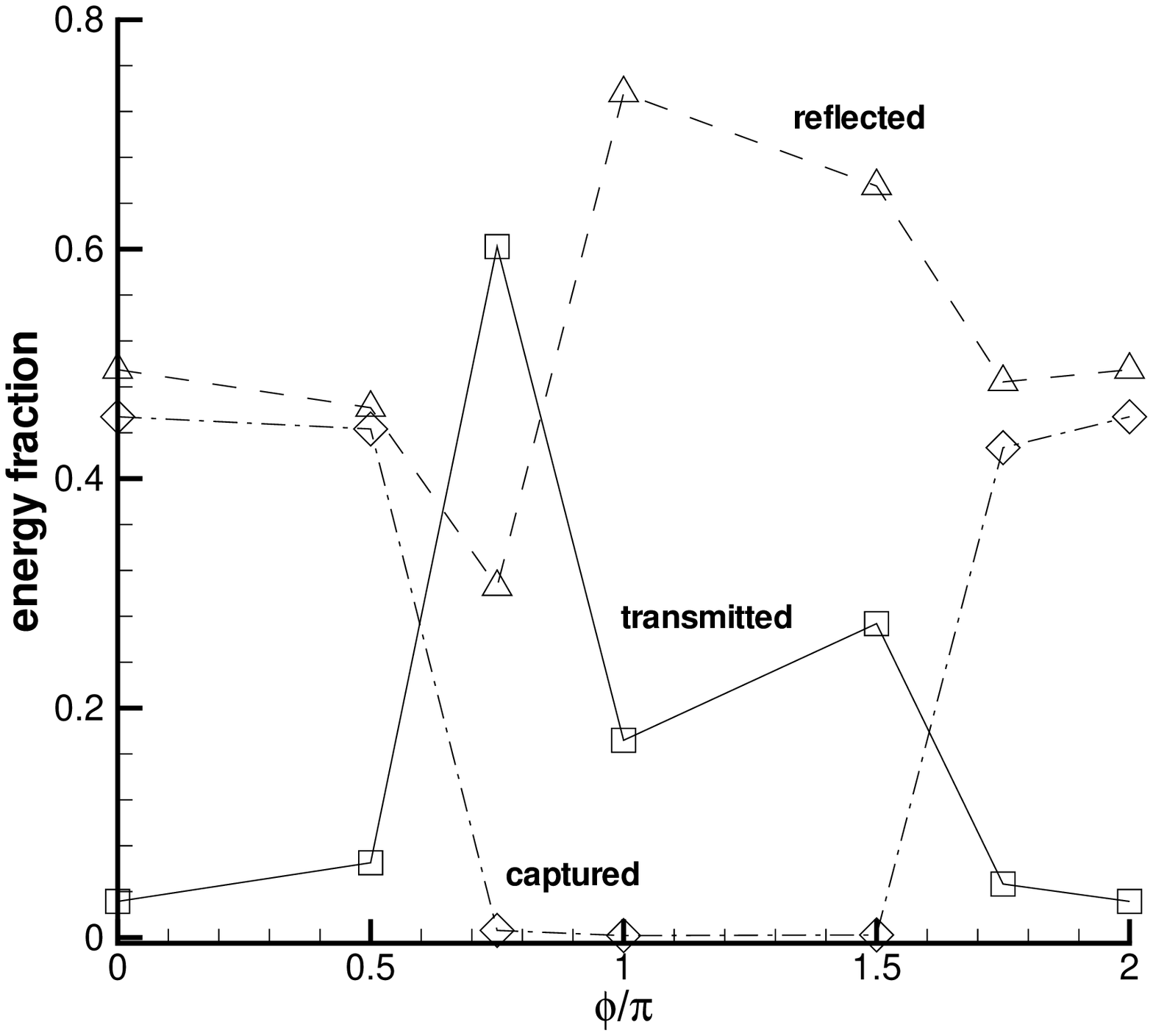}}
\caption{} \label{collision}
\end{figure}

On the contrary, the collision between the soliton originally captured by a
triplet and a free soliton is completely predictable. While the incident
soliton always bounces back, the pinned one either remains in the same state
[see an example in Fig. \ref{3Dpictures}(b)] or is kicked out in the forward
direction, depending on the velocity of the free soliton, $c$, and phase
shift $\phi $ between it and the pinned soliton at the collision moment.
Small portions of the energy are spent to form additional weak pulses, as
seen in Fig. \ref{3Dpictures}(b). The dependence of the outcome of the
collision on $c$ and $\phi $ is displayed in Fig. \ref{collision}. A sharp
transition from the rebound without depinning to the release of the pinned
soliton, in a narrow interval $0.55<|c|<0.6$, is observed in Fig. \ref%
{collision}(a). It is quite natural that the increase of the
momentum of the free soliton allows it to kick out the pinned one.
Figure \ref{collision}(b) additionally shows that the collision
effect (release of the pinned soliton) is stronger when the two
solitons are phase-shifted by $\phi \simeq \pi $ (the free
soliton, with the same momentum, kicks out the pinned one, unlike
the situation at $\phi =0$). The difference can be understood, as
in-phase solitons attract each other, while out-of-phase ones
interact with repulsion; therefore, in the former case the
interaction force acts on the pinned soliton to the right (in the
direction of the free soliton), hence it is less likely for it to
be kicked out in the opposite direction, while the repulsion
force, acting on it in the case of $\phi \simeq \pi $, makes the
eventual depinning more plausible.

It is relevant to mention that the simulations did not reveal an outcome of
the collision in the form of ``recharge", i.e., replacement of the
originally pinned soliton by the impinging one.

\section{Conclusion}

We have extended the recent analysis of the interaction of a free
soliton in the FBG\ (fiber-Bragg-grating) model with a
paired-defect set to a more complex (but still sufficiently
simple) triplet structure, formed by two repulsive defects and an
attractive one between them, with a relatively small separation
between the defects. It has been demonstrated that the triplet is
superior to other traps (including the doublet formed by
attractive and repulsive defects) in achieving the aim of the
creation of the ``standing optical soliton": the triplet allows
one to capture 2/3 of the energy of a free soliton moving at half
the light speed in the fiber. A moving intrinsically unstable
(``too heavy") soliton can also be readily captured by the
triplet, giving rise to the stationary pinned pulse. Another
advantage offered by the triplet is that the trapped soliton
quickly settles down to a stationary state, which makes the
subsequent collision of the pinned soliton with a free one
predictable: the impinging soliton always bounces back, while the
pinned one remains in the bound state or is kicked out, depending
on the velocity and relative phase of the collision. The triplet
structure, whose inner scale is on the (sub)millimeter scale, can
be readily fabricated in the FBG, and the predicted results may be
relevant to the design of soliton-based optical memory and logic
element, where solitons would serve as data bits.

\section*{Figure captions}

Figure 1. (a) Contour plots for the share of the energy of the
free soliton with velocity $c$, trapped after its collision with
the dipole (doublet) defect built as a set of the attractive and
repulsive defects with strengths $\pm \protect\kappa $. Above the
``reflection limit" boundary, the soliton as a whole bounces back,
due to the reflection from
the repulsive defect. (b) An example of the collision for $c=-0.5c$ and $%
\protect\theta =\protect\pi /2$. In this figure and below, the
defect is centered at $x=0$, and the free soliton is always
launched at $x=10$.

Figure 2. Typical examples of collisions involving a triplet. (a)
Capture of a free soliton by an empty triplet (the collision also
generates a small reflected pulse, which is poorly seen behind the
trapped soliton). (b) Bounce of a free moving soliton from the
pinned one, which was generated by
the collision shown in (a). In either case, the free soliton has amplitude $%
\protect\theta =\protect\pi /2$ and velocity $c=-0.5$. The triplet
is a set of two repulsive defects, with $\protect\kappa =-0.3$,
and an attractive one between them, with $\protect\kappa =+0.3$.

Figure 3. (a) The same as in Fig. \protect\ref{dipole}(a), but for
the capture of a free moving soliton (with amplitude
$\protect\theta =\pi /2$) by the triplet formed by two repulsive
defects with strengths $\kappa _{1}\equiv -\kappa $, and an
attractive one set between them, with strength $\kappa _{2}\equiv
\protect\kappa $. Below the ``reflection limit" boundary, the
soliton as a whole bounces back (in fact, it is a sharp border
between the capture and rebound). (b) The share of the captured
energy versus amplitude $\theta $ of the free soliton, with
velocity $c=-0.5$, and the ratio of the strengths of the repulsive
and attractive defects, if their strengths are not equal.

Figure 4. Shares of the total energy of the colliding solitons (a
free one and a soliton originally pinned by the triplet with
$\kappa _{2}=- \kappa _{1}=0.3$) which are eventually bound in the
bouncing (``reflected") pulse and in ones which stay trapped
(``captured") and move to the left, having been kicked out from
the pinned state (``transmitted"). The initial amplitude of both
solitons is $\protect\theta =\protect\pi /2$ (the bound soliton
was generated by the capture of a free one with $\protect\theta =
\pi /2$). In (a), the shares are shown versus the velocity of the
free soliton, for the collision with zero phase difference,
$\phi=0$; in (b), the same is shown versus $\protect\phi $, for
$c=-0.5$. The three shares do not add up to $100\%$, as a small
additional share is lost to radiation.

\end{document}